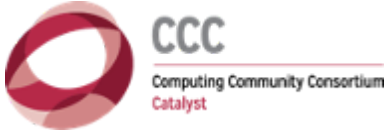

# Modernizing Data Control: Making Personal Digital Data Mutually Beneficial for Citizens and Industry

*A Computing Community Consortium (CCC) Quadrennial Paper*

*Sujata Banerjee (VMware Research), Yiling Chen (Harvard University), Kobbi Nissim (Georgetown University), David Parkes (Harvard University), Katie Siek (Indiana University Bloomington), and Lauren Wilcox (Georgia Institute of Technology)*

## Introduction

We are entering a new "data everywhere-anytime" era that pivots us from being tracked online to continuous tracking as we move through our everyday lives. We have smart devices in our homes, on our bodies, and around our communities that collect data that is used to guide decisions that have a major impact on our lives — from loans to job interviews and judicial rulings to health care interventions. We create a lot of data, but who owns that data? How is it shared? How will it be used? Who will measure the implications of data usage — including the unintended consequences? We must examine these questions to protect people — many of whom are aware that their data is being collected in exchange for services. While the average person does not have a good understanding of how the data is being used, they know that it carries risks for them and society. In addition, our data is the basis for the creation of huge economic value, hence people's data need more protections and assurances that we receive a fair value in the exchange.

When we typically talk about ownership, we talk about personal property, such as a car or house. The transaction is clear. You purchase an item and it is yours. You are responsible for deciding who can visit, borrow, share, and use it. Data ownership is less clear. When you purchase a mobile phone and a data plan, you own the physical device, but the phone company and the mobile carrier can now collect data on your usage. Data about your usage — whom you call, when, and where — also relates to the people you contact; people who were not part of the relationship you established with these companies. This situation is further exacerbated when we add in contact tracing during epidemics since being near someone may be logged by an application, phone, or wifi connection. Once individuals are thought of as participants in a network, it is harder for individuals to negotiate the value of their data, as the same or similar data can also be garnered from other individuals. This also complicates the question of ownership rights: should individuals be allowed to sell or consent for the use of their data when that use impacts others?

Likewise, when you download an app, the app itself and other sub-apps of the original app (e.g., ads within an app) also collect *explicit data* (e.g., a dietary monitoring app collects information about your diet) as well as *implicit data* (e.g., geolocation; microphone) and possibly share this with partners or sell it in "aggregate" form where data is summarized at the individual or user group level. These data ownership challenges are replicated throughout our everyday lives as we move to purchasing "smarter" (and more privacy invasive) technologies (e.g., smart watches, smart thermostats, cars) and live in a connected infrastructure (e.g., smart cities; electronic medical records with personal portals). Data goes beyond the data we generate, but also to our preferences, which can be aggregated to create our profile, potentially more detailed than our knowledge of ourselves, with the pretext of enabling personalization.

Although some people may believe they own their data — an impression enhanced by mechanisms companies created so people can download their data — in reality, the problem of understanding the myriad ways in which data is collected, shared, and used, and the consequences of these uses is so complex that only a few people want to manage their data themselves. Furthermore, much of the value in the data cannot be extracted by individuals alone, as it lies in the connections and insights garnered from (1) one's own personal data (is your fitness improving? Is your home more energy efficient than the average home of this size?) and (2) one's relationship with larger groups (demographic group voting blocks; friend network influence on purchasing). But sometimes these insights have unintended consequences for the person generating the data, especially in terms of loss of privacy, unfairness, inappropriate inferences, information bias, manipulation, and discrimination. There are also societal impacts, such as effects on speech freedoms, political manipulation, and amplified harms to weakened and underrepresented communities. For instance, insurance companies provide people with the ability to upload health tracker data, without disclosure on possible uses that could impact medical coverage. Smarthome data can be subpoenaed and authorities can issue search warrants to collect pacemaker data that can be used against the owner in the court of law. Group connections can lead to targeted media campaigns to suppress votes. To this end, we look at major questions that policymakers should ask and things to consider when addressing these data ownership concerns.

**What are the implications for using data?**

If people cannot own their data, then data management morphs into *access control*. With access control, multiple parties can *access* data, but care must be taken to ensure the **data will not adversely impact those who generated the data.** Although researchers warn that "requiring exclusivity [not freely sharing data] without a good reason for doing so would unnecessarily limit what can be done with the data," we do not have a good understanding of **what kinds of unintended consequences can result from different levels of data access.** We must invest in research to better understand these consequences and enable third parties to investigate these consequences in the broader population.

Part of ownership is also the right to reproduce data. For example, if I purchase a movie, I can make a copy of it for my own personal use. Similarly, with data about ourselves, we should be able to make a copy of it — especially with the risk of data being locked down by ransomware. But, we still must ask **who should benefit and who may be harmed** when entities copy data?

### How do we quantify the value of data?

Data is valuable: it generates billions of dollars a year in personalized advertisements and boosts the efficiency of advertising marketplaces. When data is employed for beneficial purposes, researchers can use it to make transformational contributions to society — in the natural sciences, the social sciences, and computing. These contributions in turn drive innovations in the private sector, especially in medical, financial, and business, where people can benefit through personalized, efficient, and effective transactions. When considering proposals that use peoples' data, one should consider **what data will be needed and what will the benefit be for individuals, society, and the entities involved in the creation of knowledge enabled by that data.**

These innovations rely on analyzing people's data — sometimes sharing data between multiple third party entities. Firms can utilize this data for personalized, behavioral marketing, thus further capitalizing on consumer data. It has also been suggested that firms can use consumer data to gain a competitive advantage over smaller businesses. Data sharing between entities is technically challenging because systems may not be interoperable and usage policies may conflict, even within the same company. At the same time, the sharing of even anonymized data can leave the possibility of re-identification: most US citizens can be identified using 3-15 demographic attributes (e.g., ZIP code, gender, and date of birth). Ultimately, this means that current anonymization practices are insufficient, especially in relation to the California Consumer Privacy Act (CCPA) and European General Data Protection Regulation (GDPR). Thus, policymakers should carefully consider **how and when one's data** — even in aggregate — **is shared between third parties** since the data can be used adversely. A hypothetical example would be if a federally funded research team identifies those at risk for heart disease in their state and provides their data to the public as part of their data management plan. A third party could utilize the data to identify who is at risk to guide decisions about employment or life insurance coverage.

### How can we control data?

When we are considering access and control to data, we should consider **who the real users are**. In some instances, our current notions of legal responsibility (e.g., a cognizant adult) may not be safe or realistic (e.g., minors may need agency of their data - especially in medical contexts, instead of their guardians). In addition, we need **mechanisms where data ownership can easily be shared between trusted people** (e.g., from a minor to a young adult; an independent adult who suffers cognitive decline and needs assistance).

A new direction for data control is suggested by *data co-ops*, which are an approach that looks to leverage collective negotiation power through access to legal, economic, and technical tools that enable individuals and groups to express their preferences about the use of their collective information. Data co-ops preserve values such as privacy, fairness, and security, while enabling users to negotiate, oversee, and enforce the terms of how their data is utilized by others (including any compensation). Data co-ops provide mechanisms for the redistribution of value back to co-op members*.* A data co-op relies on policy and technological tools. Further research is needed to **develop interventions towards improving the balance between individuals and data platforms** and, importantly, a **rigorous understanding of how these interventions can be put together to better achieve societal goals**, in particular, how to design tools that integrate policy, technological, and economic components.

Other possibilities include *federated data stores*, where people would keep their data on their individual devices and decide what they want to share, with whom, and when. In this situation, "trades" that have been implicit can be made explicit. For example, targeted ads in return for access to preferred digital media. More research is needed to better **understand how well people will be able to negotiate these kinds of trades while preserving their privacy**.

These types of data control would require "mechanisms to channel, constrain, and facilitate the flow of data" to help people understand the value, benefits, and possible consequences of data sharing. In these types of scenarios, people would be empowered to participate in the monetization and utilization of their data. **We need a deeper understanding of how to make people aware** of their participation in these systems and **considerations for how industries could fairly negotiate with people.** Although the CCPA and GDPR are good steps forward for peoples' data use, we need an approach that better balances all competing interests within the United States diverse cultural and economic interests.

**How can we protect peoples' data?**

In the US, we are currently in a market-driven approach to data ownership and use, where everyday people have little power. For example, I can opt out of giving my data to a company (e.g., Google), but the company still has my data based on my everyday interactions with it (e.g., Google has everything from one's web searches; internet browsing on Chrome; everyday and business interactions on their suite of apps; and cell phone uses with an Android phone unless one explicitly selects being anonymous). Currently, people opt-in to terms of service so they can benefit from services in the short term, however we need **policies to protect people over the long term** and throughout their lives, especially in times of crisis.

The California Consumer Privacy Act provides people the right to know what data is being collected; who has access to the data; and the ability to say they do not want their data sold. Essentially, there is **transparency, but not forced compliance.** More research must be done on **how data transparency should be presented to diverse, representative populations** of the United States and what sanctions should be introduced at the policy level for non-compliance.

Overall, we see a need for people to be able to control the use of their data; extract value from their data; and understand their exposure to harms. This data can be narrowly defined as an individual's personal data or a group's data (e.g., a family; ethnic group; social network). More work must be done by policymakers and researchers because, while data in the large is valuable, the value of any individual's data is low and impossible to capture alone. As a result, we do not have a market where individuals choose how to gain value from their data, and we do not have good tools for assigning value to individual data. We need to work towards having mechanisms to protect both *outgoing information* (e.g., control of information provided by people; how the information is used; and assurances against adversarial uses) and *incoming information* (e.g., regulation of data-driven services provided to people).

*This material is based upon work supported by the National Science Foundation under Grant No. 1734706. Any opinions, findings, and conclusions or recommendations expressed in this material are those of the authors and do not necessarily reflect the views of the National Science Foundation.*

*For citation use: Banerjee S., Chen Y., Nissim K., Parkes D., Siek K. & Wilcox L. (2020) Modernizing Data Control: Making Personal Digital Data Mutually Beneficial for Citizens and Industry.* [https://cra.org/ccc/resources/ccc-led-whitepapers/#2020-quadrennial-papers](https://cra.org/ccc/resources/ccc-led-whitepapers/#2020-quadrennial-papers)